\documentstyle[prl,aps,psfig,epsf,twocolumn]{revtex}
\begin{document}
\draft
\twocolumn[\hsize\textwidth\columnwidth\hsize\csname@twocolumnfalse%
\endcsname

%\preprint{}

\title{Tunneling between Luttinger liquids}
\author{Paul Fendley$^1$ and Chetan Nayak$^2$}
\address{$^1$Physics Department, University of Virginia,
Charlottesville, VA 22904--4714\\
$^2$Physics Department, University of California Los Angeles,
Los Angeles, CA 90095--1547}

\date{August 16, 2000}
\maketitle
\vskip 0.5 cm
\begin{abstract}

We consider the problem of tunneling between spinless
$1D$ Luttinger liquid chains.
We show how to map the problem onto the $4$-state
chiral clock model together with a free boson for the total charge
mode.  We use a variety of results, some of them exact,
from the integrable chiral Potts
model and the study of commensurate/incommensurate transitions to
deduce the physics of coupled Luttinger liquids.
For those {\it intra}chain interaction strengths
for which {\it inter}chain tunneling is relevant, we find that
it can lead to the formation of symmetric and antisymmetric
bands with split Fermi surfaces,
depending on the relative strengths of interchain
tunneling and {\it inter}chain interactions.
With split Fermi surfaces, interchain transport is coherent.
It is not possible to have two gapless Fermi surfaces with
the same Fermi momentum when interchain tunneling
is relevant. However,  interchain interactions
can drive the formation of a gap, in which
case interchain transport will be incoherent.
We comment on the possible relevance of our results
to $c$-axis transport in high-$T_c$
superconductors.
\end{abstract}

\vspace{1 cm}

\vskip -0.4 truein
\vspace{0.5 cm}
]
\narrowtext
%\newpage

\section{Introduction}

The problem of tunneling between $1D$ Luttinger
liquids has been a focus of interest and controversy.
Much of the interest stems from the unusual
phenomenology of the cuprate superconductors.
The normal state of these materials exhibits
insulating ($d{\rho_c}/dT<0$) or, at best, very poor
$c$-axis transport coexisting with metallic
$ab$-plane transport \cite{Ong94}. Moreover,
angle-resolved photoemission \cite{Loeser96,Campuzano99}
reveals a Fermi surface with no dispersion along the
$c$-axis, despite the fact that the hopping matrix
elements are quite substantial between the two
planes of a Bi2212 bi-layer. Apparently, single-electron
motion along the $c$-axis is very strongly suppressed
while motion in the $ab$-plane is not.

It has been suggested that this peculiar situation is
characteristic of a putative non-Fermi liquid normal
state. $1D$ Luttinger liquids are the best-understood non-Fermi
liquid metals; as a result, they are a natural testing
ground for such ideas. Indeed, it has been suggested
\cite{Clarke94} that the type of behavior seen
in the cuprates occurs when two
$1D$ Luttinger liquids are coupled by weak single-electron
hopping between the $1D$ chains.

When intrachain interactions
are {\it extremely} strong,
interchain hopping between
$1D$ Luttinger liquids is
irrelevant in the renormalization group sense.
Extremely strong means much stronger than can be reached in
the infinite-$U$ Hubbard model,
and so strong that the fermion spectral function has
no singularity whatsoever at the Fermi surface.
In such a case, it is possible to
have metallic behavior within a chain but
insulating transport between chains.
For weak and even moderately strong
intrachain interaction
strengths, however, interchain hopping is a relevant perturbation.
The natural assumption is that this leads to metallic
interchain conduction.

Many analyses
\cite{Finkelstein93,2leg,VZ,Balents96,multichain,Schulz96}
diagonalize the interchain hopping term first,
forming symmetric and antisymmetric bands with
displaced Fermi surfaces in the $2$-chain
case. Interactions are then treated as perturbations.
Of course, if these perturbations could be treated exactly, the resulting
solution would be the same as that which would be
obtained by diagonalizing intrachain interactions first
and then solving interchain hopping (which we will do
non-perturbatively and, in one interesting limit, exactly,
as we describe below).
However, this need not be true for approximate solutions;
since the approximate solutions of
\cite{Finkelstein93,2leg,VZ,Balents96,multichain,Schulz96}
assume $c$-axis dispersion (i.e.\ dispersion
in the direction perpendicular to the chains, which
we'll call the $c$-axis) by taking the existence of
symmetric and antisymmetric bands
as a starting point, it is not surprising
that they do not find the scenario of
\cite{Clarke94}.

Consequently, it has not been possible
to rule out claims \cite{Clarke94}
that interchain hopping,
though relevant, does not lead to {\it coherent}
transport (we discuss the possible meaning of this
below) and, thus, does not lead to $c$-axis dispersion
or metallic interchain transport. These claims have been
supported by appealing to the spin-boson problem --
in which a two-level system is coupled to a bath
of bosons -- which has a parameter regime in which
tunneling between the two levels is relevant, but
there are no coherent oscillations \cite{Leggett87,Chakravarty95}.
The authors of \cite{Clarke94} have claimed that there is a
similar range of intrachain interaction strengths
for which interchain hopping between $1D$
Luttinger liquids is relevant but
does not lead to metallic interchain transport.

To resolve this issue, we must address whether
there is $c$-axis dispersion
and metallic interchain transport.
If the minimum energy for adding an electron
occurs at two different Fermi surfaces, $\pm {k_F^s}$ and
$\pm {k_F^a}$, corresponding to the symmetric and
antisymmetric bands (i.e.\ if there are two different wavevectors
for the minimum energy excitations which are
even and odd under the
$Z_2$ symmetry which exchanges the two Luttinger
liquids), then there is
$c$-axis dispersion. If, in addition,
this minimum energy is zero,
then interchain transport will be metallic.
In this case, various long-wavelength
power-law correlation functions
will oscillate at wavevector $\pm \left({k_F^s}-{k_F^a}\right)$.
According to \cite{Clarke94}, there is no gap,
but this splitting does not occur:
${k_F^s}={k_F^a}$.

However, the problem is complicated somewhat by the fact that
interchain {\it interactions} can drive the
formation of a gap. When this occurs, it
is possible to have $c$-axis dispersion without
metallic interchain transport. In this case,
the minimum energy for adding an electron
occurs at two different `Fermi surfaces', $\pm {k_F^s}$ and
$\pm {k_F^a}$, but this energy is non-zero.
The opening of such a gap would not explain
the absence of a bi-layer splitting in photoemission
studies of the cuprates \cite{Loeser96,Campuzano99}
since it occurs `on top' of the bi-layer splitting.
On the other hand, it is also possible for the
gapped phase to show no $c$-axis dispersion,
so that the minimum energy (which is non-zero)
for adding an electron occurs at a single `Fermi surface'.
Clearly, there is no phase transition
between these two limits of the
gapped phase, but they embody a qualitative
distinction which is important
for photoemission in the cuprates.
The purpose of this paper is to determine
which of these possibilities is realized.
We do this by diagonalizing intrachain interactions first
and then treating interchain tunneling and interchain interactions
by non-perturbative methods, some of which are exact.

The main technical difficulty which has prevented the
resolution of the problem of tunneling between Luttinger liquids
is that tunneling is a chiral perturbation -- i.e.\
the interchain tunneling
term transforms non-trivially under the $O(2)$ rotation
group of $1+1$-dimensional Euclidean (in imaginary-time formalism)
spacetime. Physically, the tunneling operator is chiral because
it mixes right-moving fermions in one chain with
right-moving fermions in the other, and similarly
for left-moving fermions; it does not mix right-movers
with left-movers. This stands in contrast to the case of a non-chiral
perturbation -- which mixes right-movers
with left-movers -- for which the $1+1$-dimensional
sine-Gordon model is the archetype. As we know from this exactly
soluble models, non-chiral perturbations, if relevant,
usually lead to the formation of a gap. Chiral
perturbations are poorly understood, however.
At the free fermion point, interchain tunneling
is a chiral perturbation which is soluble
because it is quadratic in fermion operators; it
leads to the formation of symmetric and antisymmetric
bands with Fermi surfaces which are shifted from
the unperturbed Fermi surface. However, the behavior at this
trivial point may not be generic. Until now,
it has not been possible to appeal to a
non-trivial exactly soluble model in which the
fate of a relevant chiral perturbation is understood.

In this paper, we consider spinless $1D$ Luttinger liquids 
coupled by tunneling.
We believe that spin is an inessential
complication for the issue at hand; the claims of
\cite{Clarke94} apply to spinful and spinless Luttinger
liquids, as do possible analogies to the spin-boson problem.  We map
this model to a statistical mechanical model
known as the generalized $4$-state chiral clock model.  The phase
structure of this model is fairly well understood \cite{Ostlund,dN}.
This chiral clock model has a disordered phase
and both commensurate and
incommensurate ordered phases. As we discuss below,
the incommensurate phase corresponds to an
electronic state with $c$-axis dispersion and metallic
interchain transport.
This picture receives further support from the consideration
of a particular limit in which the model becomes
the integrable chiral Potts model which,
according to exact results \cite{McCoy,Cardy},
is in an incommensurate phase.

In the interesting regime in
which interchain tunneling is relevant,
there are two possible phases.
Depending on the relative strength
of interchain interactions and interchain hopping,
the system can either be in a gapless incommensurate
phase or a gapped phase. In the incommensurate
gapless phase, the coupled Luttinger chains
exhibit $c$-axis dispersion and
metallic interchain transport despite strong
intrachain non-Fermi liquid properties.
It is not possible for
the system to be gapless and commensurate;
in other words, there is no gapless phase with two
identical Fermi surfaces if the tunneling matrix
element is non-vanishing. Hence, some of the
claims of \cite{Clarke94} are incorrect.
The gapped phase, on the other hand, may or may not have displaced
Fermi surfaces. Thus, it is possible that the mysterious $c$-axis
physics of the high-$T_c$ cuprates can be explained by a simple
analogy with the $1D$ Luttinger liquid, but only in
the somewhat extreme parameter regimes in which
$c$-axis tunneling is irrelevant or a gap forms which is
strong enough to preclude $c$-axis dispersion.

In section 2, we introduce the model of coupled Luttinger liquids and
give its bosonized formulation.
In section 3, we review the results for tunneling
between Fermi liquids, and see how incommensurability arises there.
In section 4, we map the Luttinger problem to a generalized 4-state
chiral clock model. In section 5, we use results from the chiral clock
model to understand the physics of coupled Luttinger liquids.

\section{The Model}

The Luttinger liquid is a model of interacting gapless fermions in one
spatial dimension.  At the microscopic level, it might be
described by the lattice Hamiltonian
\begin{eqnarray}
{H_0} = -t {\sum_{<i,j>}}\left({c^{\dagger}_{i}}{c^{}_{j}} + {\rm h.c.}
\right) + {\sum_{i,j}}V(|i-j|) {n_{i}}{n_{j}}
\end{eqnarray}
where $i,j$ label sites along the $1D$ chains, $c_i$ annihilates
a fermion at site $i$, and
${n_{i}}={c^{\dagger}_{i}}{c^{}_{i}}$.
In the long-wavelength limit, the corresponding
Luttinger liquid action is:
\begin{eqnarray}
{S_0}  =  {\int}\! dx\int \! dt\biggl(
{c^{\dagger}_{R,L}}\left({\partial_t} \pm
v {\partial_x}\right){c^{}_{R,L}}\,-
u\,{c^{\dagger}_{R}}{c^{}_{R}}\,
{c^{\dagger}_{L}}{c^{}_{L}}\biggr)
\end{eqnarray}
where the fermion creation operator is
\begin{eqnarray}
{c}(x) = {e^{-i{k_F}x}}\,\,{c_{R}}(x)\,\,+
\,\,{e^{i{k_F}x}}\,\,{c_{L}}(x).
\label{creation}
\end{eqnarray}
As is well known (see \cite{Stone} and references therein),
the Luttinger model
may be rewritten in terms of
a single free boson with action
\begin{equation}
{S_0}  =  {\int}dx \int dt\,
\frac{1}{8\pi}\,\left({\left({\partial_t}{\varphi}\right)^2}
- {v^2}{\left({\partial_x}{\varphi}\right)^2}\right)
\label{eqn:szero}
\end{equation}
The field, $\varphi$, is taken to be an angular variable satisfying
the periodicity condition ${\varphi}\equiv {\varphi} + 2\pi r$. Here
$r=\sqrt{g}$, where $g$ is determined by the interaction strength,
$g=\sqrt{(v-u)/(v+u)}$. The coupling $g<1$ for repulsive interactions,
$g>1$ for attractive interactions. Our results apply to both cases
because the Hamiltonian is invariant under the duality symmetry $g\to
1/g$. In the long-wavelength limit, the theory is
`Lorentz invariant', with $v$ playing the role of the speed
of light.

The operators of the theory are defined in terms of chiral
fields obeying 
$\left({\partial_t}\pm v{\partial_x}\right){\varphi_{R,L}}=0$.
Then ${\varphi} = {\varphi_{L}} + {\varphi_{R}}$, while the dual field
$\widetilde{\varphi}= {\varphi_{L}} - {\varphi_{R}}$. It is convenient
to label the operators by their ``electric'' charge $m$ and 
``magnetic'' charge $n$, defined as
\begin{equation}
\{m,n\} \equiv  \exp\left(i{\frac{m}{2r} \varphi + inr\widetilde\varphi}
\right)
\label{elecmag}
\end{equation}
The left and right scaling dimensions of this operator are then
\begin{equation}
(h_L,h_R)=\left(\frac{1}{2}\left(\frac{m}{2r} + nr\right)^2,
\frac{1}{2}\left(\frac{m}{2r} - nr\right)^2 \right).
\end{equation}
This means that when $r=\sqrt{g}$,
the overall scaling dimension $x=h_L+h_R$ and Lorentz spin $s=h_L-h_R$
of the operator $\{m,n\}$ are
$$x= \frac{m^2}{4g} + n^2g\, ,\qquad \qquad s= mn.$$
The electron operator in particular is the $\{1,1/2\}$ operator, namely
\begin{eqnarray}
{c_{L}} = {e^{\frac{i}{2}
\left(\left(\sqrt{g}+1/\sqrt{g}\right){\varphi_{L}} -
\left(\sqrt{g}-1/\sqrt{g}\right){\varphi_{R}}\right)}}
\end{eqnarray}
Then $c_L^\dagger=\{-1,-1/2\}$, while $c_R=\{1,-1/2\}$ and
$c_R^\dagger=\{-1,1/2\}$.  The electron operators have Lorentz spin
$\pm 1/2$, meaning that they have non-trivial properties under Lorentz
transformations (i.e.\ rotations in the in 1+1 dimensional
spacetime). Operators with non-zero Lorentz spin are called {\it
chiral}. Adding a chiral operator to the action breaks Lorentz invariance.

In this paper, we consider the effects of coupling two
identical
spinless Luttinger liquids. 
There are two types of couplings which will concern us here,
interchain hopping and interchain interactions.
Interchain hopping takes the form:
\begin{eqnarray}
{H_{c}} = -{t_\perp} {\sum_{i}}
\left({c^\dagger_{i,1}}{c^{}_{i,2}} + {\rm h.c.}
\right)
\end{eqnarray}
while interchain interactions take the form:
\begin{eqnarray}
{H_{V_\perp}} = {\sum_{i,j}}{V_\perp}(|i-j|) {n_{i,1}}{n_{j,2}}
\end{eqnarray}
Here, ${c^{}_{i,1}}$ annihilates an electron at site
$i$ of chain 1 while ${c^{}_{i,2}}$ annihilates an electron at site
$i$ of chain 2; ${n_{i,1}}$ and ${n_{j,2}}$ have
corresponding definitions.

In the long-wavelength limit, we can rewrite the model in terms of two
free bosons $\varphi_1$ and $\varphi_2$. However, the interactions
depend only on the difference of the two, so we define the linear
combinations
\begin{equation}
{\varphi_\pm}=\frac{1}{\sqrt{2}}\left({\varphi_1}\pm{\varphi_2}\right)
\end{equation}
Then interchain hopping takes the form:
\begin{eqnarray}
{{\cal L}_c}  &=& -{t^R_\perp}\,
{e^{\frac{i}{\sqrt{2}}
\left[\left(\sqrt{g}+1/\sqrt{g}\right)
{\varphi_{\!-R}} - \left(\sqrt{g}-1/\sqrt{g}\right)
{\varphi_{\!-L}}\right]}} \,+\,\,c.c.\cr
& & -{t^L_\perp}\,
{e^{\frac{i}{\sqrt{2}}
\left[\left(\sqrt{g}+1/\sqrt{g}\right)
{\varphi_{\!-L}} - \left(\sqrt{g}-1/\sqrt{g}\right)
{\varphi_{\!-R}}\right]}} \,+\,\,c.c.
\end{eqnarray}
In the decoupled theory, the dimensions just add, so
this operator has dimension and spin twice as
large as the fermion, namely $x=(g+1/g)/2$ and Lorentz spin $\pm 1$.
Thus this theory has chiral operators in the action
and breaks Lorentz invariance.
In a model of coupled chains which is parity-invariant,
${t^L_\perp}={t^R_\perp}$. However, it will be convenient to allow
these two hopping parameters to be independent so as to separate
various effects. In particular, the case ${t^L_\perp}=0$,
${t^R_\perp}\neq 0$ will be of special interest.  Interchain
interactions take the form:
\begin{eqnarray}
{{\cal L}_{V_\perp}}  &=& {V^{}_{2k_F}}\,
\cos(\sqrt{2g}{\tilde \varphi}_{\!-}) 
\,+\, {\cal J} \cos(\sqrt{2/g}\varphi_{\!-}).
\end{eqnarray}
The first is a pure interaction term of dimension $2g$: no charge is
transferred between the two liquids. The second is a Josephson
tunneling or pair hopping term;
it transfers a pair of electrons from one chain to
the other. Even if we set these terms to zero by hand (and
there is no reason to do so), second-order perturbation theory in
the interchain hopping induces them.
This model is invariant under the electric-magnetic duality $g\to 1/g$:
under this symmetry
the pair hopping and interchain interactions are swapped.
We could also add zero-momentum density-density and
current-current interactions
of the form $\partial_{x,t}{\varphi_1}\,\partial_{x,t}{\varphi_2}$,
but they will not change the results described here: they only
renormalize the coupling $g$ and the Fermi velocity $v$, so we absorb
them in redefinitions of the couplings.

Thus the interactions of two coupled Luttinger liquids can be
described by a single boson $\varphi_{\!-}$; the boson $\varphi_+$ is
free and describes the total charge, which is
unaffected by the interchain coupling.  
It is useful to
rewrite the interaction terms in the action in terms of electric and magnetic
charges with a radius $r=\sqrt{g/2}$. The
interchain hopping terms are of the form (\ref{elecmag}), namely
$\{\pm 1,\pm 1\}_-$, where the subscript indicates the extra
$\sqrt{2}$ in the relation of the radius to $g$. The $V_{2k_F}$ term
is $\{0,\pm 2\}_-$, while the ${\cal J}$ term is $\{\pm 2,0\}_-$.

The lowest-order RG equations for these perturbations are:
\begin{eqnarray}
\label{eqn:rg_pert}
\frac{d {t^{R,L}_\perp}}{d\ell} &=& \left(2 -
\frac{1}{2}\left(g+\frac{1}{g}\right)\right){t^{R,L}_\perp}
\,+ \,O\left({t^{R,L}_\perp}\,{t^R_\perp}{t^L_\perp}\right)
\cr
\frac{d {{\cal J}}}{d\ell} &=& \left(2- 2/g\right){\cal J}\,+ 
\,O\left({t^R_\perp}{t^L_\perp}\right)\cr
\frac{d {V_{2k_F}}}{d\ell} &=& \left(2 - 2g\right){V^{}_{2k_F}}
\,+ \,O\left({t^R_\perp}{t^L_\perp}\right)
\end{eqnarray}
Several important features of the model are apparent in these
equations. 

%\begin{list}
\medskip
\noindent
(1) For $g\ne 1$, either the pair-hopping or the $2k_F$
interaction is relevant, and the other one is irrelevant. Thus when
the electrons interact, there is always a relevant non-chiral term in
the action.

\medskip\noindent
(2)  For $1>g>2-\sqrt{3}=.268\dots$,  ${t^{R}_\perp}$, ${t^{L}_\perp}$
and ${V_{2k_F}}$ are relevant. For $g<2-\sqrt{3}$
tunneling between Luttinger liquids is irrelevant.
For $g<2-\sqrt{3}$, the fermion operator has dimension
$1$, so its spectral function has no singularity
at the Fermi surface (for $2-\sqrt{3}<g<1$, there is at least
a power-law divergence; for $g=1$, i.e.\ free fermions,
a $\delta$-function singularity); hence, this is the
limit of {\it extremely} strong interactions.

\medskip\noindent
(3) If ${t^L_\perp}=0$ and ${V_{2k_F}}={\cal J}=0$ but ${t^R_\perp}\neq 0$,
then all the higher-order terms in the RG equation for ${t^R_\perp}$ vanish:
higher powers of ${t^R_\perp}$ are of higher Lorentz spin
and thus cannot renormalize $t^R_\perp$.
This is in stark contrast to the pair-hopping
and $2k_F$ interaction terms which do renormalize
themselves.

\medskip
\noindent
Our aim in this
paper is to discuss the fate of the system under the flow of these
relevant perturbations.

\section{Weak-Coupling Physics}

We first consider the free fermion limit, $g=1$. If we set
${V^{}_{2k_F}}={\cal J}=0$, then we simply have two Fermi liquid chains
coupled by single-fermion hopping. This special case was discussed in
detail in \cite{pryadko}. The hopping operators are of dimension
$(1,0)$ and $(0,1)$. As a result of the interchain hopping, symmetric
and antisymmetric bands are formed with Fermi surfaces displaced from
$k_F$. For the purpose of comparison with $g\neq 1$, we describe this
in a little more detail.

The action is
\begin{eqnarray}
{S_0}  &=&  {\sum_{I=1,2}}{\int}dx\int dt\,\biggl(
{c^{\dagger}_{I R,L}}\left({\partial_t} \pm
v {\partial_x}\right){c^{}_{I R,L}}\,+
\cr & & {\hskip 1 cm}
- {t^R_\perp}\,{c^{\dagger}_{1 R}}{c^{}_{2 R}} + c.c
- {t^L_\perp}\,{c^{\dagger}_{1 L}}{c^{}_{2 L}} + c.c
\biggr)
\end{eqnarray}
where $I=1,2$ specifies the chain.
We diagonalize this by forming symmetric and antisymmetric bands in
terms of
\begin{eqnarray}
{c^{}_{s,a\, R,L}}(x) &=& \left({c^{}_{1R,L}}(x) \pm  {c^{}_{2R,L}}(x)
\right)\,{e^{\frac{i}{2v}{t^{R}_\perp}\,(vt \pm x)}}
\end{eqnarray}
The action is simply:
\begin{eqnarray}
\label{eqn:symm_anti}
S  &=&  \int dx \int dt\,\biggl(
{c^{\dagger}_{s R,L}}\left({\partial_t} \pm
{v_s} {\partial_x}\right){c^{}_{s R,L}}\,+\cr
& & {\hskip 2.1 cm} {c^{\dagger}_{a R,L}}\left({\partial_t} \pm
{v_a} {\partial_x}\right){c^{}_{a R,L}}\biggr).
\end{eqnarray}
If the fermion spectra are truly strictly linear for
${t^{R,L}_\perp}=0$, then ${v_s}={v_a}=v$;
otherwise, these velocities will be shifted.

It is instructive to re-cast these steps in
the bosonic representation, which is the
most useful representation for $g\neq 1$.
Then,
\begin{eqnarray}
S  &=&  {\int}dx \int dt\,\,
\frac{1}{8\pi}\,\left({\left({\partial_t}{\varphi_\pm}\right)^2}
- {v^2}{\left({\partial_x}{\varphi_\pm}\right)^2}\right)\,\,-\cr
& & {\int}\,dx\,\int\,dt\,\left(
{t^R_\perp}\,
{e^{i{\varphi_{\!-R}}\sqrt{2}}}\,
+\, {t^L_\perp}\,{e^{i{\varphi_{\!-L}}\sqrt{2}}}\,+\,c.c.\right)
\end{eqnarray}

The $SU(2)$ symmetry which interchanges the two
uncoupled chains can be used
to transform this into a more tractable form. Performing
a rotation generated by the unitary operator
\begin{eqnarray}
U=\exp\left(\frac{\pi}{2}\left({e^{i{\varphi_{\!-R}}\sqrt{2}}}
- {e^{-i{\varphi_{\!-R}}\sqrt{2}}} + L\leftrightarrow R\right)\right)
\end{eqnarray}
we transform the action into:
\begin{eqnarray}
\label{eqn:hopping_rot}
S  &=&  {\int}dx \int dt\,\,
\frac{1}{8\pi}\,\left({\left({\partial_t}{{\varphi'}_\pm}\right)^2}
- {v^2}{\left({\partial_x}{{\varphi'}_\pm}\right)^2}\right)\,\,-\cr
& & \int dx \int dt\,\left(
{t^R_\perp}\,{\partial_-}{{\varphi'}_{\!-R}} +
{t^L_\perp}\,{\partial_+}{{\varphi'}_{\!-L}}\right)
\end{eqnarray}
where ${\varphi'}$ is the rotated field,
${\partial_\pm}={\partial_t}\pm v{\partial_x}$ and,
for simplicity, we have assumed that ${t^{R,L}_\perp}$ are real.
The second line of (\ref{eqn:hopping_rot}) can be
eliminated by making the change of variables:
${{\varphi'}_{\!-R,L}}\rightarrow {{\varphi'}_{\!-R,L}} +
\frac{1}{2v}{t^{R,L}_\perp}(vt\mp x)$.
As a result, various correlations functions
of the unshifted variables will be spatially
modulated:
\begin{equation}
\label{eqn:trans_corr}
\left\langle{e^{i{{\varphi'}_{\!-R}}(x)/\sqrt{2}}}
{e^{-i{{\varphi'}_{\!-R}}(0)/\sqrt{2}}}\right\rangle
= \frac{1}{x^2}\,\,{e^{\frac{i}{v}{t^{R}_\perp}\,x}}
\end{equation}
The same is true of the unprimed fields.
This spatial modulation of a correlation function
is characteristic of an incommensurate
phase, and can be taken as the definition of such
a phase \cite{dN}.
This behavior still occurs even if
${t^{R}_\perp}\neq{t^{L}_\perp}$. In particular,
if ${t^{R}_\perp}\neq 0$, but ${t^{L}_\perp}=0$,
the modulation occurs in
 the right-moving sector, but not the left-moving one.
The relation to the fermionic representation is:
\begin{eqnarray}
{c^{}_{s,a\, R,L}}(x) &=& 
{e^{\frac{i}{\sqrt{2}}{\varphi_{\!+R}}}}\,
{e^{\pm\frac{i}{\sqrt{2}}{{\varphi'}_{\!-R}}\pm
\eta \frac{i}{2v}{t^{R,L}_\perp} x}}
\end{eqnarray}
so comparing with (\ref{creation}) we see how in an incommensurate
phase the fermi surfaces are shifted.

To summarize, the hallmark of $c$-axis dispersion
and metallic $c$-axis transport
is the oscillation of the power-law correlation
functions of ${\varphi_{\!-}}$, as exemplified
by eq. (\ref{eqn:trans_corr}). This is equivalent to the
statement that gapless fermionic excitations
occur at two different Fermi surfaces.

Let us now turn on weak interactions and move
away from the free fermion point, but remaining
in the weak-coupling regime. In this regime,
it makes sense to consider both intra- and interchain
interactions as a perturbation of the system in
which interchain hopping has already been solved
(\ref{eqn:symm_anti}). These interactions take the form:
\begin{eqnarray}
\label{eqn:symm_anti_int}
{S_{\rm int}} &=&  \int dx \int dt\,
\biggl[{u_s}\,{c^{\dagger}_{s R}}{c^{}_{s R}}\,{c^{\dagger}_{s L}}{c^{}_{s L}}
\,+\cr & & {\hskip 2.1 cm}
{u_a}\,{c^{\dagger}_{a R}}{c^{}_{a R}}\,{c^{\dagger}_{a L}}{c^{}_{a L}}\cr
& &+\,{u_x}\,\left({c^{\dagger}_{s R}}{c^{}_{s R}}\,
{c^{\dagger}_{a L}}{c^{}_{s L}}
+ {c^{\dagger}_{a R}}{c^{}_{a R}}\,
{c^{\dagger}_{s L}}{c^{}_{s L}}\right)\cr
& &+\, {u_t}\,\left({c^{\dagger}_{s R}}{c^{\dagger}_{s L}}\,
{c^{}_{a R}}{c^{}_{a L}} +
{c^{\dagger}_{a R}}{c^{\dagger}_{a L}}\,
{c^{}_{s R}}{c^{}_{s L}}
\right)\biggr]
\end{eqnarray}

The renormalization group equations for
these interactions read:
\begin{eqnarray}
\label{eqn:weak_sa_RG}
\frac{d {u_{s,a}}}{d\ell} &=& \frac{{v_s}+{v_a}}{2 v_{s,a}}\,\,
{u_t^2}
\cr
\frac{d {u_x}}{d\ell} &=& - {u_t^2}\cr
\frac{d {u_t}}{d\ell} &=& - \left[\frac{{v_s}+{v_a}}{2 v_s}\,{u_s} +
\frac{{v_s}+{v_a}}{2 v_a}\,{u_a} - 2{u_x}\right]\,{u_t}
\end{eqnarray}

For ${v_s}={v_a}$, these
are the Kosterlitz-Thouless equations
for $u_t$ and ${u_s} + {u_a} - 2{u_x}$.
In the limit of small $u_t$, the system is
described by two gapless decoupled
Luttinger liquids (corresponding to
the symmetric and antisymmetric bands)
if ${u_s} + {u_a} - 2{u_x}>0$. If,
in the same limit, we instead have
${u_s} + {u_a} - 2{u_x}<0$, then
a gap is opened. The action (\ref{eqn:symm_anti_int})
is Lorentz invariant in this limit,
so this gap cannot affect the splitting
between the Fermi momenta of the symmetric
and antisymmetric bands. The lowest energy
single-fermion excitations will be at
$\approx {k_F} \pm \frac{1}{2v}{t^{R}_\perp}$.
When we move away from ${v_s}={v_a}$
(which will certainly happen if the interaction strengths
are different in the two bands), we can no longer appeal to
Lorentz invariance, and it is possible that the
symmetric and antisymmetric bands are not split in momentum.
In the limit of very large ${u_s} + {u_a} + 2{u_x}$,
the entire approach breaks down because interchain tunneling
between the two chains is irrelevant.
In this limit, symmetric and antisymmetric
bands do not form, and $c$-axis transport is
insulating. In subsequent sections,
we address the question of whether
this can happen in the strong-coupling
limit even when interchain tunneling
is still relevant.

\section{The chiral clock model}

In this section we introduce a lattice model which, in the continuum
limit, is described by the above coupled Luttinger liquids, including
the chiral operators. In the next section we will utilize known
results on the incommensurate phases of this lattice model to extract
physics for our problem.

The lattice model is a generalized chiral clock model, with
nearest-neighbor interactions on a square lattice. It is written in
terms of discrete ``clock'' variables $2a\pi/N$, where $a$ is an
integer ranging from $0$ to $N-1$.  The interaction energy between
neighbors of values $a$ and $b$ is required to have a ${\bf Z}_n$
symmetry and, thus, is
of the form ${\cal E}(a,b)={\cal E}(a-b)={\cal
E}(a-b+N)$. The general action is
\begin{eqnarray}
\label{latticeaction}
S=-\sum_{x,y}\sum_{j=1}^{N-1} 
&&\Big( 
K_j \cos\left[\frac{2\pi }{N} (jD_xa - \Delta_j)\right]\cr
&&
\left. +\widehat{K_j}
\cos\left[\frac{2\pi}{N}(jD_ya - \widehat{\Delta}_j)\right]
\right)
\label{clockaction}
\end{eqnarray}
where $D$ is the lattice derivative, namely $D_xa\equiv
a(x+1,y)-a(x,y)$ and $D_ya\equiv a(x,y+1)-a(x,y)$.  We can obtain the
physics we want by taking
\begin{eqnarray}
\nonumber
K_j=\widehat{K}_j \quad&&\quad K_{N-j}=K_j\cr
\Delta_1=-\Delta_{N-1}=\Delta \quad&&\quad
\widehat{\Delta}_1=-\widehat{\Delta}_{N-1}=\widehat{\Delta}\cr
\Delta_j=&&\widehat{\Delta}_j=0\quad\hbox{otherwise}
\end{eqnarray}
We will also take $K_j$ to be real, but $\Delta$ and
$\widehat{\Delta}$ need not be real. Usually, clock models
have $K_j=0$ for $j>1$, but it will prove very
useful to relax this condition.
This model was originally introduced in \cite{Ostlund}, which
discussed the special case $K_1=1/T$ and $K_j=0$ for $j>1$, and
$\widehat{\Delta}=0$.  When $\Delta\ne 0$, the model is chiral.  For
$0<\Delta\le 1/2$, it is energetically more favorable for $D_xa$
to be $1$ than for it to be $-1$: $a$ tends to rise (cyclically)
as $x$ moves to the right. The fact that at low temperature there are
two possibilities (no rise or a rise of $+1$) leads to the
existence of an incommensurate phase, to which we will return in the
next section.

For couplings close to a critical point, one expects the continuum
physics of the chiral clock model to be described by a field theory. For
$\Delta=\widehat{\Delta}=0$, a number of critical points are well
understood. For $N$=3, there is a single critical point at
$K_1=[\ln(\sqrt{3}+1)]/3$, the $S_3$-symmetric 3-state Potts model
critical point. The corresponding field theory is a conformal field
theory with central charge $c=4/5$. For $N\ge 5$, there is a critical
``Villain'' line. This line appears by arguments familiar from the
Coulomb gas approach to the Kosterlitz-Thouless transition in the XY
model. Namely, one assumes that the discrete clock variable $a(x,y)$
can be replaced by a bosonic field $\phi(x,y)$ taking continuous
values: the effect of the original discreteness is a potential $\cos
(N\phi)$ (the operator $\{N,0\}$ in the notation of section 2). When
$N\ge 5$ there is a regime in which both vortices (operators $\{0,n\}$
for $n$ integer) and the $\cos(N\phi)$ potential are irrelevant
\cite{Jose}. In this regime the free boson fixed-line is therefore
stable. In the full parameter space of the generalized clock
model with no chirality, there is a region which flows to this stable
fixed line. However, there is an additional fixed point in this
parameter space, known as the Fateev-Zamolodchikov or parafermion
fixed point
\cite{FZ,para}. This unstable fixed point is at the edge of the region
of stability of the Villain line, and is at a self-dual value of the
couplings. The parafermion conformal field theory describing this
fixed point has central charge $c=2(N-1)/(N+2)$ while the Villain line
has central charge $c=1$.  As $N$ gets higher, the structure gets more
elaborate, but the parafermion and Villain points are the only ones we
need to discuss here; see \cite{Dorey} for a thorough discussion of
the phase structure of the ${\bf Z}_5$ and ${\bf Z}_6$ models.

The case of interest for the coupled Luttinger liquids turns out to be
$N=4$. The existence of the Villain line follows directly from the
fact that the (non-chiral) ${\bf Z}_4$ clock model is equivalent to the
Ashkin-Teller model. The Ashkin-Teller model is two coupled Ising
models, with $s,\sigma=\pm 1$ on every site related to $a$ via
$$e^{i\pi a/2}=\frac{s+i\sigma}{1+i}$$
so at $\Delta=\widehat{\Delta} =0$ the action becomes
\begin{equation}
\label{AT}
S=-\sum_{<ij>}\left({K_1}(\sigma_i\sigma_j + s_is_j) + K_2
\sigma_i\sigma_j s_i s_j\right).$$
\end{equation}
The original chiral clock model is at
$K_2=0$, where the two Ising models decouple.
On the self-dual line $\sinh(2K_1)=\exp(-2K_2)$, the Ashkin-Teller
model is, in turn, equivalent to a special case of the lattice 8-vertex
model which is not only integrable but critical for $K_2\le \ln(3)/4$
\cite{Baxter}.  The continuum field theory description of this
critical line is precisely a free boson  (or to be precise, its orbifold
\cite{SKY}). That the Ashkin-Teller model
is equivalent to a free boson should not be surprising.
The two-dimensional Ising model is equivalent to a free Majorana
(real) fermion, so two Ising models are equivalent to a Dirac
fermion. In the continuum theory, the four-fermion coupling is the only
way to add interactions without moving off the critical line. The $N=4$
parafermion critical point is a special point on this line, namely
$K_2=(\ln(\sqrt{2}+1))/4$. The critical point of the four-state Potts
model (which has a full $S_4$ symmetry) is on the self-dual line at
$K_2=K_1=(\ln 3)/4$.

Now that we have found a critical line of the lattice model and the
corresponding field theory, we need to find the operators in the
field theory which are relevant perturbations. The thermal perturbation is
not chiral: it corresponds to moving off the self-dual line while
keeping $\Delta=\widehat\Delta=0$. This model is not integrable on the
lattice, but the dimension of the thermal operator is known by combining the
exact solution at $T=T_c$
\cite{Baxter} with standard scaling assumptions and a Coulomb-gas analysis
\cite{Nien}.  We can put this result in our notation for free
bosons by defining the radius $r_{AT}$ via
$$\cos\left(\frac{\pi}{8r_{AT}^2}\right)\equiv\frac{1}{2}(1-e^{4K_2})$$
The thermal operator then is $\{0,\pm 2\}_{AT}$, with left and right
dimensions ($2r_{AT}^2,2r_{AT}^2$).  At the $K_2=0$ decoupled point,
$r_{AT}=1/2$, so the thermal perturbation has dimensions $(1/2,1/2)$:
it corresponds to giving a mass to the Ising fermions. At the
parafermion fixed point, the thermal operator is of dimension
$(1/3,1/3)$.

We find the operators in the long-wavelength
field theory which couple to
$\Delta$ and $\widehat\Delta$ by following the
argument in \cite{Cardy}. When $\Delta$ and $\widehat\Delta$ are
small, the linear term in (\ref{latticeaction}) can be rewritten
as
\begin{eqnarray}
\nonumber
K_1 \left(\Delta \sin[\frac{2\pi j}{N}(D_x a)]
+\widehat\Delta 
\sin[\frac{2\pi j}{N}(D_ya)]\right).
\end{eqnarray}
As one would expect from difference operators, $\sin D_x f$ and $\sin
D_y f$ transform into each other under $90^o$ lattice rotations. In
other words, these operators are not invariant under Lorentz
transformations: they transform like vectors. In the continuum theory,
they should therefore have Lorentz spin $\pm 1+4n$, for some
integer $n$. The chiral perturbation must correspond to adding to the
action operators with Lorentz spin $\pm 1$ (operators with higher spin
are irrelevant). We can identify these operators unambiguously by
using one other fact about this perturbation: second-order
contributions of the chiral terms renormalize the original thermal
perturbation, because $2\sin^2x= 1+\cos 2x$. Thus the operator product
of the chiral operator (Lorentz spin $1$) with the antichiral operator
(Lorentz spin $-1$) must contain the thermal operator.  Under operator
products, the electric and magnetic charges merely add. This means
that the chiral operators must be $\{1,1\}_{AT}$ and $\{-1,-1\}_{AT}$,
while the antichiral operators are $\{1,-1\}_{AT}$ and
$\{1,-1\}_{AT}$.

For example, at the point $K_2=0$ where the Ising models decouple, the
interaction between Ising spins connected by a bond in the $x$ direction is
$$K_1 \cos \Delta (\sigma_i\sigma_{i+1} + s_is_{i+1})
+ K_1 \sin \Delta (\sigma_is_{i+1} - s_i\sigma_{i+1}).
$$
The second term is thus proportional to $\sigma D_x s -
sD_x\sigma$. The continuum field theory therefore 
has action
\begin{eqnarray}
\nonumber
S=S_{I^2} &+& \Delta_L \int
\left(\sigma(x,t) \partial_L s(x,t) -
s(x,t) \partial_L \sigma(x,t)\right)\cr
&+&\Delta_R
\int \left(\sigma(x,t) \partial_R s(x,t) -  
s(x,t) \partial_R \sigma(x,t)\right)\cr
\end{eqnarray}
where $S_{I^2}$ is the action of two decoupled Ising models, and
$\Delta_{L,R}=(\Delta \pm i\widehat{\Delta})$. We have continued to
real time $t$ via $y= it$.  The operator product of two magnetization
operators in the Ising model indeed contains the thermal operator, so
the second-order contribution of this term indeed obeys our above
condition.  Since the magnetization fields $\sigma,s$ of the Ising
models have dimensions $(1/16,1/16)$, the operator coupling to
$\Delta_L$ has dimensions $(9/8,1/8)$, while the operator coupling to
$\Delta_R$ has dimensions $(1/8,9/8)$.  These operators are not total
derivatives, so in the bosonic language they do indeed correspond to
the operators $\{\pm 1,\pm 1\}_{AT}$ and $\{\pm 1,\mp 1\}_{AT}$ at the
decoupling radius $r_{AT}=1/2$.

The action of the continuum field theory describing the $4$-state
chiral clock model near its critical line is therefore found by adding
all the different perturbations to the free boson action $S_0$:
\begin{eqnarray}
\nonumber
&&S=S_{0}+(T-T_c) \int (\{0,2\} + \{0,-2\})\cr
&&+\Delta_L \int (\{1,1\} + \{-1,-1\}) +\Delta_R \int (\{1,-1\}
+ \{-1,1\}).
\end{eqnarray}
All the operators are defined
with radius $r_{AT}$.  This is the action of our coupled Luttinger
liquid written in terms of $\varphi_{\!-}$ !  One needs to make the
identification $t_\perp^{L,R}=\Delta_{L,R}$, $T-T_c = V_{2k_F}$, and
$r_{AT}=\sqrt{g/2}$. The critical line and hence the
description is valid for
$e^{4 K_2} < 3$, which corresponds to $1/8 < r^2_{AT} < 3/8$.  Thus the
coupled Luttinger liquid is equivalent to the continuum limit of
the generalized chiral clock
model if $1/4 < g < 3/4$.
For the dual regime $4>g>4/3$, one instead identifies
$r_{AT}=\sqrt{1/2g}$ and $T-T_c = {\cal J}$.

\section{Coupled Luttinger Liquid Physics from the Chiral Clock
Model}

The chiral clock models were the subject of quite a bit of study a few
decades ago; a review can be found in \cite{dN}. The four-state model
with $K_2\ne 0$ was discussed in \cite{AYP}. When
$\widehat\Delta=0$, the four-state model is believed to have a phase
diagram which looks like Fig. 1. 

\begin{figure}
%{\includegraphics[scale=0.6]{zero.ps}}
\centerline{\epsfxsize=3.6in\epsffile{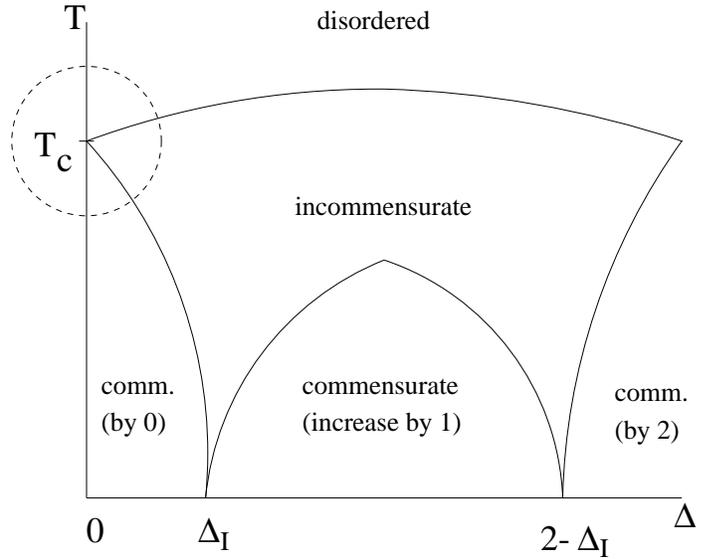}}
\bigskip\bigskip
\caption{The phase diagram of the four-state chiral clock model
at a fixed value of $K_2/K_1$ and $\widehat{\Delta}=0$; $\Delta_I$ is given by
(\ref{deltaI}). The field theory
description should be valid in some region around $T=T_c$, $\Delta=0$.}
\end{figure}

This picture is for some fixed value
of $K_2/K_1$ (fixed $g$ in the Luttinger problem). The critical point
at $\Delta=0$ is the corresponding critical point on the Ashkin-Teller
self-dual line; changing the temperature corresponds to changing $K_2$
and $K_1$ while keeping $K_2/K_1$ constant. This phase diagram should
be valid for $2-\sqrt{3}<g<3/4$. Since everything is invariant under
the transformations $g\to 1/g$, the models are also equivalent for
$4/3<g<2+\sqrt{3}$. For $g<2-\sqrt{3}$, the chiral
perturbation is irrelevant, so there is no incommensurate phase.
The other bound of the map, $g=3/4$, is the value
$K_2 =-\infty$. As shown in section 3, there is definitely an
incommensurate phase at the free fermion point $g=1$.  As we will
argue below, we expect three types of behavior.  For $g<2-\sqrt{3}$
and its dual, the model is always commensurate.  For
$2-\sqrt{3}<g<3/4$ and its dual, both phases are possible, depending
on the relative strength of the interchain interactions and the
interchain hopping.  Because as one brings $g$ closer to $1$ the
hopping operator becomes more relevant and the non-chiral operator
less relevant, our guess is that the incommensurate phase takes a
larger portion of the phase diagram as $g\to 3/4$.  For $4/3>g>3/4$,
we argue that the commensurate phase vanishes. These possibilities are
summarized in figure 2. Explaining these phase diagrams is the purpose
of this section.

\begin{figure}
%{\includegraphics[scale=0.6]{zero.ps}}
\centerline{\epsfxsize=2.5in\epsffile{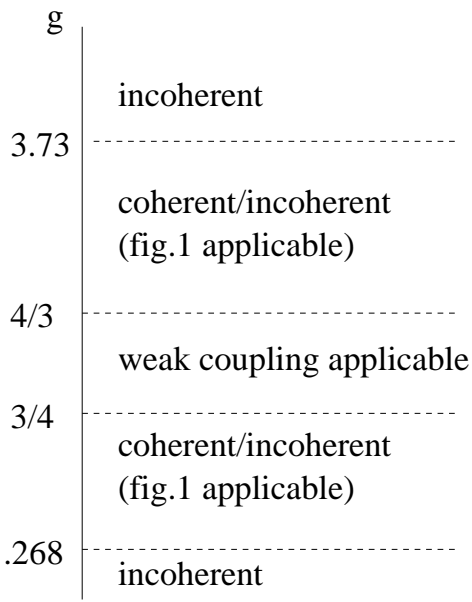}}
\bigskip\bigskip
\caption{The phases of the coupled Luttinger liquids
as a function of the intrachain interaction
strength $g$.}
\end{figure}

The evidence for the existence of an
incommensurate phase in the chiral clock model
derives from several arguments.  In the parity-invariant model
($\widehat{\Delta}=0$), the main evidence comes from looking at the
lattice model close to zero temperature ($K_1\to\infty$,
$K_2\to\infty$ with $K_2/K_1$ fixed). At zero temperature, the system
is completely ordered, except at special points
at which  there are two
lowest-energy states. In the four-state model, this happens at
$\Delta=\Delta_{I}\le 1/2$, where \cite{AYP}
\begin{equation}
\frac{K_2}{K_1}=
\sqrt{2}\sin\left[\frac{\pi}{2}\left(\Delta_{I}-\frac{1}{2}\right)\right].$$
\label{deltaI}
\end{equation}
At zero temperature, if $\Delta<\Delta_I$, $a$ takes the same value
at every site; but if $\Delta_I<\Delta<2-\Delta_I$, $D_x a = 1$,
i.e.\ $a$ increases by one (cyclically)
every site as one moves to the right. These are
commensurate phases.  At $\Delta=\Delta_{I}$, it is equally
favorable energetically for
$D_x a$ to be $0$ or $1$. Thus there are a huge number of ground
states, characteristic of an incommensurate phase.  At
$\Delta=\Delta_{I}$ (while still $\widehat\Delta=0$ and $T=0$), the
system forms stripes with constant $a$ in the $y$-direction.
\cite{Ostlund}. The values of $a$ increases (cyclically)
from stripe to stripe as $x$ increases.

We can extend the argument of \cite{Ostlund} to $K_2\ne 0$ to show
that there is an incommensurate phase at small but nonzero values of
temperature. At non-zero temperature, the stripe boundaries are no
longer necessarily strictly parallel to the $y$-axis: they can develop
kinks.  The partition function can be computed at low temperatures if
one neglects stripe creation (i.e.\ stripes which do not exist for all
values of $y$). Defining $\rho$ to be the number of stripes per unit
length, one finds
\begin{equation}
\cos(\pi \rho)=  e^{2K_1+2K_2} \left(K_1\sqrt{2}
\sin\left[\frac{\pi}{2}\left(\frac{1}{2}-\Delta\right)\right]+K_2\right)
\label{region}
\end{equation}
When the magnitude of the right-hand side is greater than one, the
phase is ordered and commensurate. In the range of $K_1$ and $\Delta$
where this magnitude is less than 1, the density of stripes varies
continuously with the parameters. This is the incommensurate phase.

The low-temperature approximation thus accounts for the bottom part of
the phase diagram.
We now consider the evidence supporting
the upper part of the phase diagram of figure 1.
First of all, it is useful to consider what happens for clock
models for $N>4$. As noted before, there is a region of
the phase diagram in which vortices
(which correspond in the low-temperature picture
to stripe-creation operators)
and the $\cos N\phi$ clock operator are irrelevant. The chiral operators
coupling to $\Delta$ still remain relevant. Thus
even at $\Delta=0$ there is a range of temperatures with a phase
intermediate between commensurate and disordered, in the manner of
\cite{HN}. It is natural to assume that as one turns on $\Delta$, this
intermediate phase turns into the incommensurate phase. Thus the phase
diagram for $N>4$ looks like figure 1, except that the
incommensurate/disordered transition line hits the $y$-axis above the
commensurate/incommensurate line. For $N=4$, there is
a particular point on the $y$-axis at which
the vortex and the clock operators become (naively)
marginal, while, as we have seen, the chiral operators remain
relevant.  This results in the picture of figure 1, in which the
disorder/incommensurate and commensurate/incommensurate lines
hit the $y$-axis at the same point. The system can be
tuned across the disorder/incommensurate boundary by
varying the chiral and non-chiral
perturbations, i.e.\ by changing the interchain tunneling
and interaction strengths in the coupled Luttinger liquid model.

As $g$ increases to $3/4$, the size of the incommensurate region
increases because the exponent in (\ref{region}) decreases.  In fact,
as $g\to 3/4$, $K_2\to -\infty$ while $K_2/K_1\to -1$. This means
effectively that the critical point at $T=T_c$ is moving towards the
$T=0$ axis. In addition, the incommensurate value $\Delta_I$ also is
moving to $0$. Moreover, at $g=3/4$, $2(K_1+K_2)=\ln 2$ on the
self-dual line so the exponential factor in (\ref{region}) goes away.
Thus the commensurate phase in the field theory is being pushed out.
Therefore, we believe that in the low-temperature phase at $g=3/4$ the
model is gapless even at $\Delta=0$ (and that the low-energy fixed
point turns out to be the {\it three}-state Potts critical point).
There is no longer a commensurate phase in the field theory.  We
believe that for $4/3>g>3/4$, only the incommensurate and disordered
phases are possible, which dovetails with our analysis of the
weak-coupling RG equations (\ref{eqn:weak_sa_RG}). These equations are
of Kosterlitz-Thouless type, and there is no evidence for any
commensurate phase. 

Thus we see that the $g\approx 1$ weak-coupling picture should persist until
$g$ is lowered to $3/4$. For $2-\sqrt{3}<g<3/4$, the phase diagram in
Fig.\ 1 is applicable.  In the continuum picture, this phase diagram
results from the interplay between the chiral and non-chiral
operators.  At $T=T_c$ and $\Delta=0$, the coefficient of the thermal
(interchain interaction) operator is tuned to be zero, and we have a
critical point. Taking $\Delta>0$ while keeping $|V_{2k_F}|$ small
then results in the incommensurate phase. Thus we believe that the
phase structure is determined by the relative strengths of the
tunneling and the interactions.  (We will present additional evidence
for this below.)  We expect the size of the incommensurate phase to
get bigger as $g$ increases to $3/4$, because of the low-temperature
argument above, and because the chiral tunneling operator is getting
more relevant while the non-chiral interactions are getting less
relevant.

The incommensurate phase here is believed
to be gapless. The main evidence for this comes from a
closely-related model: the sine-Gordon field theory in a background
field coupling to the charge \cite{Sch,HBB}.  This model clearly has a
phase diagram similar to that of figure 1.
The commensurate phase comes when the
$\cos(N\phi)$ term is relevant and the background field is small. For
large enough background field, it is energetically favorable to have
kinks in the ground state, even though they are gapped. This results
in the incommensurate phase, which clearly is gapless, because
increasing the background field continuously increases the number of
kinks in the ground state.

Note that the disordered phase can have minimum energy
excitations which are commensurate or incommensurate.
The former case occurs in the limit in which the
non-chiral perturbation is much larger than the chiral
perturbation. The latter was discussed in
the context of the weak-coupling analysis at
the end of section III. Both of these possibilities
lie within the disordered phase of the chiral clock
model and are, hence, adiabatically connectable.

Note also that our results predict the existence of a Lifshitz point
(i.e.\ a point at which the commensurate, incommensurate, and disordered
phases all coexist) in the chiral clock model
for $1/4 < K_2/K_1 < 2-\sqrt{3}=.268\dots$.
In this regime, the chiral operator is irrelevant
near the $\Delta=0$ critical point, so the disordered/commensurate
transition line should extend out for finite $\Delta$. However, for
$g>1/4$ there is still an incommensurate phase at low-enough
temperature. The phase diagram should look almost like figure 1,
except there is a direct phase transition between
the disordered and commensurate phases for $\Delta$
small. The phase boundary between the disordered and commensurate phases
ends at a Lifshitz point, beyond which these two
phases are separated by the incommensurate phase.

Strong support for the above picture comes from an integrable model,
usually called the chiral Potts model \cite{McCoy}. This model
comprises a special subspace of the couplings in the general clock
Hamiltonian (\ref{clockaction}).  The name is somewhat misleading: the
integrable model is not $S_N$ symmetric
but only ${\bf Z}_N$ symmetric.
When there is no chiral interaction, the
integrable model reduces to the ${\bf Z}_N$ Fateev-Zamolodchikov
parafermion critical point (there is no Potts critical point for
$N>4$).

Unfortunately, except at the critical point
the couplings in the integrable model do not include the value
$\widehat{\Delta}=0$ in which we are most interested. However, it does
include a very interesting special case, corresponding to $\Delta=\pm
i\widehat{\Delta}$.  Because of the $i$, the
Boltzmann weights are not real, but the Hamiltonian in real time is
Hermitian.  In the language of the coupled Luttinger model, this
corresponds to setting the tunneling at one of the Fermi points, say
$\Delta_R=t^R_\perp=0$. With this condition, the model remains
self-dual even away from the critical line \cite{HKN}.  The thermal
operator is not self-dual, so it cannot be in the continuum action
either \cite{Cardy}. Thus the field theory of the integrable model on
this line is equivalent to that of Luttinger liquids at $g=1/3$ (the
parafermion value) coupled by a single chiral tunneling operator
\cite{Cardy}. This allows us to isolate the physics of the
chiral operator (with coefficient $\Delta_L=t^L_\perp$) from that
of the thermal ($V_{2k_F}$) perturbation.

At a special value of $\Delta_L$ known as the superintegrable
point, the exact excitation energies can be computed \cite{Roan}.  One
finds that level crossing has occurred: there are states with $Z_N$
charge (not to be confused with the electric charge
of our Luttinger liquids)
which have energy below that of the original ground state,
which is chargeless. Thus the ground state at non-zero
temperature contains a mixture of these $Z_N$ charged states, just as in
the sine-Gordon model at large enough background fields.  Thus, at the
superintegrable point on the self-dual line, the model is gapless and
incommensurate. Moreover, this phase persists even when a small
non-zero $\Delta_R$ is allowed.

One very interesting exact result is that the exponents are the same
at the superintegrable point and the parafermion critical point, once
the anisotropy is properly accounted for. This is a strong indication
that the model is gapless and incommensurate as one tunes $\Delta$
from $0$ to the superintegrable point, and in fact all along the
self-dual line. All other exact results support this assertion as well
\cite{Roan}.  The continuum language also strongly supports this
picture \cite{Cardy}. Because the critical point is perturbed only by
a single chiral operator, the thermal operator is not induced in
perturbation theory. The only effect of renormalization is to scale
the coefficient of the chiral operator, and not create a gap. This is
also implied by the analysis of the sine-Gordon model in a background
field \cite{HBB}, which indicates that there is no Lifshitz point in
the three-state model.

Thus the integrable model provides strong evidence that adding a
single chiral operator to the action results in a gapless
incommensurate phase. This is in perfect harmony with the picture that
adding chiral operators to the action tends to cause an incommensurate
phase, while adding non-chiral ones causes commensurability.  A
transition occurs as the relative couplings are varied. Unfortunately,
we do not know how to determine the exact location of the transition
line (except at low temperatures in the lattice model). Nevertheless,
we believe that all available evidence agrees with the phase diagram of
figure 1.

\section{Conclusions}

Our main conclusions are

\medskip
\noindent
(1) For $g=1/3$,
interchain tunneling
alone (which can be disentangled from interactions
by taking ${t^L_\perp}={V_{2k_F}}=0$ but ${t^R_\perp}\neq 0$)
leads to the formation of symmetric and antisymmetric
bands with displaced Fermi surfaces -- i.e.
$c$-axis dispersion and metallic interchain
transport.

\medskip
\noindent
(2) Interchain interactions
which couple the $2k_F$ density oscillations of the
two chains suppress $c$-axis dispersion and transport.
As a result, for $3/4>g>2-\sqrt{3}$ (and the mathematically equivalent
dual case $4/3<g<2+\sqrt{3}$),
the relative strengths of interchain tunneling
and interchain interactions determine the phase boundary
between a state with $c$-axis dispersion and
metallic $c$-axis transport and a state with a
partially gapped spectrum (the total charge mode
remains gapless) which exhibits insulating $c$-axis transport.
As interchain interactions
are increased within the gapped phase,
the $c$-axis dispersion smoothly
disappears; no phase boundary is crossed as
the minimum energy excitations move to the same
point in momentum space.

\medskip
\noindent
(3) There is no fully gapless phase without $c$-axis dispersion
and metallic interchain transport, contrary to
the claim of \cite{Clarke94}.

\medskip

These results were
drawn by mapping the problem of coupled Luttinger liquids onto
the $4$-state chiral clock model.
The phase diagram of the latter model can be obtained
from an analysis of the low- and high-temperature expansions
of this model, together with an analysis of the
$Z_4$ parafermion critical point of the non-chiral
limiting case of the model. Result (2)
follows from this phase diagram. Note, in particular,
that this phase diagram has a single
gapless phase: the incommensurate phase.
Hence, it is not possible to have a completely
gapless phase with two identical Fermi surfaces
unless $g<2-\sqrt{3}$, in which case interchain
tunneling is irrelevant (this case is beyond the regime
of validity of the mapping to the $4$-state chiral
clock model). Result (1)
was obtained from exact solutions of the
$4$-state chiral Potts model (a special case
of the general $4$-state chiral clock model).
These conclusions are consistent with
earlier analytical 
\cite{Finkelstein93,multichain,2leg,VZ,Balents96,Emery99,Schulz96}
and numerical \cite{Poil} studies.
They contradict some of the claims of
\cite{Clarke94}, although
we do find that it is possible for $c$-axis transport to be
suppressed in a regime in which the tunneling is still relevant.

These results suggest the following lessons for the interlayer
physics of the high-$T_c$ superconductors.  In-plane non-Fermi liquid
behavior alone is not likely to be enough to frustrate single-electron
hopping between copper-oxide planes in high-$T_c$
superconductors. However, the formation of a gap (possibly with
interlayer correlations) could have this effect,
thereby leading to a state which
is metallic within the planes but does not
show $c$-axis dispersion.

%\bibliography{../udw/corr}

\begin{references}


\bibitem{Ong94}
N.~P. Ong,  in {\em Non-Fermi-liquid aspects of charge transport in the cuprate
  superconductors} , pp.\ 221--4, International
  Conference on Materials and Mechanisms of Superconductivity - High
  Temperature Superconductors IV Grenoble, France 5-9 July 1994 Physica C, and
  references therein.

\bibitem{Loeser96}
A.~G. Loeser, Z.~X. Shen, D.~S. Dessau, D.~S. Marshall, C.~H. Park, P.
  Fournier, and A. Kapitulnik, Science {\bf 273},  325  (1996).

\bibitem{Campuzano99}
J.~C. Campuzano, H. Ding, M.~R. Norman, and M. Randeria,  in {\em Destruction
  of the Fermi surface in underdoped cuprates}, pp.\
  517--21, international Conference on Strongly Correlated Electron Systems
  Paris, France 15-18 July 1998, Physica B.

\bibitem{Clarke94}
D. Clarke, S. Strong, and P. Anderson, Phys. Rev. Lett. {\bf 72},  3218
  (1994); Phys. Rev. Lett. {\bf 74},  4499  (1995).

\bibitem{Finkelstein93}
A. M. Finkel'stein  and A.I. Larkin, {Phys. Rev. B} {\bf 47}, 10461 (1993)

\bibitem{2leg}  M.~Fabrizio, {Phys. Rev. B} {\bf 48}, 15838 (1993);
D.~V.~Khveshchenko and T.~M.~Rice, {Phys. Rev. B} {\bf 50}, 252 (1994). 

\bibitem{VZ} C.~M.~Varma and A.~Zawadowski, {\prb} {\bf 32}, 7399 (1985).

\bibitem{Schulz96}
H.J. Schulz, Phys. Rev. B {\bf 53}, 2959 (1996), cond-mat/9412098.

\bibitem{Balents96}
L. Balents and M. Fisher, Phys. Rev. B {\bf 53}, 12133 (1996),
cond-mat/9503045.

\bibitem{multichain}  For a review, see E.~Dagotto and T.~M.~Rice, {\it %
Science} {\bf 271}, 618 (1996), cond-mat/9509181.


\bibitem{Emery99}
 V.~J. Emery, S.~A. Kivelson, O. Zachar,
Phys. Rev. B {\bf 59}, 15641 (1999), cond-mat/9810155.

\bibitem{Leggett87}
A.J. Leggett, S.~Chakravarty, A. Dorsey, M.P.A. Fisher, A. Garg, and W.
  Zwerger, Rev. Mod. Phys. {\bf 59}, 1  (1987).


\bibitem{Chakravarty95}
S. Chakravarty and J. Rudnick, Phys. Rev. Lett. {\bf 75},  501  (1995).

\bibitem{Ostlund} S.~Ostlund, Phys. Rev. {\bf B24} (1991) 398.

\bibitem{dN} M. den Nijs, in {\it Phase Transitions and Critical Phenomena},
ed. by C. Domb and J. Lebowitz, vol. 12, (Academic Press, 1988)

\bibitem{McCoy} H. Au-Yang, B. McCoy, J. Perk, S. Tang, and M.L. Yan,
Phys. Lett. {\bf A123} (1987) 219; B. McCoy, J. Perk, S. Tan and C.H. Sah,
Phys. Lett {\bf A125} (1987) 9; R. Baxter, J. Perk and H. Au-Yang,
Phys. Lett. {\bf A128} (1988) 138

\bibitem{Cardy} J. Cardy, Nucl. Phys. {\bf B389} (1993) 577, hep-th/9210002

\bibitem{Stone} see e.g.\ M. Stone, {\it Bosonization} (World Scientific, 1994)

\bibitem{pryadko} J. Naud, L. Pryadko, S. Sondhi, 
Nucl.Phys. {\bf B565} (2000) 572, cond-mat/9908188

\bibitem{Jose} J. Jose, L. Kadanoff, S. Kirkpatrick and D. Nelson, 
Phys. Rev. {\bf B16} (1977) 1217

\bibitem{FZ} V.A.~Fateev and A.B.~Zamolodchikov, Phys. Lett. {\bf 92A}
(1982) 37

\bibitem{para} A.B.~Zamolodchikov and V.A.~Fateev, Sov. Phys. JETP {\bf 62}
(1985) 215

\bibitem{Dorey}  P. Dorey, R. Tateo and K. Thompson, Nucl. Phys. {\bf B470}
 (1996) 317, hep-th/9601123; P. Dorey, P. Provero, R. Tateo, S. Vinti, 
J.Phys. {\bf A32} (1999) L151, hep-th/9810202 

\bibitem{Baxter} R. Baxter, {\it Exactly Solved Models in Statistical
Mechanics} (Academic Press, 1982)

\bibitem{SKY} S.K. Yang, Nucl. Phys. {\bf B285} (1987) 183

\bibitem{Nien}B. Nienhuis in {\it Phase Transitions and Critical Phenomena},
ed. by C. Domb and J. Lebowitz, vol. 11, (Academic Press, 1987)


\bibitem{AYP} H. Au-Yang and J.H.H. Perk, Physica {\bf A228} (1996) 78.

\bibitem{HN} D. Nelson and B. Halperin, Phys. Rev. {\bf B19} (1979) 2457

\bibitem{Sch} H. Schulz, Phys. Rev. {\bf B22} (1980) 5274
\bibitem{HBB}
F.D.M. Haldane, P. Bak and T. Bohr, Phys. Rev. {\bf B28} (1983) 2743

\bibitem{HKN} S. Howes, L. Kadanoff and M. den Nijs, Nucl. Phys. {\bf B215}
(1983) 169 

\bibitem{Roan} G. Albertini and B. McCoy, Nucl. Phys. {\bf B350} (1991) 745;
B. McCoy and S. Roan, Phys. Lett. {\bf A150} (1990) 347; in {\it Special
Functions} ICM-90 (Springer-Verlag, 1991).

\bibitem{Poil} D. Poilblanc, H. Endres, F. Mila, M. G. Zacher, S. Capponi, 
W. Hanke, Phys. Rev. {\bf B54} (1996) 10261, cond-mat/9605106;
S. Capponi, D. Poilblanc, F. Mila, Phys. Rev. {\bf B54} (1996) 17547
cond-mat/9608004

\end{references}
%\bibliographystyle{prsty}

\bigskip
We would like to thank Alexei Tsvelik for organizing the Newton
Institute conference where this work was initiated. We thank Duncan
Haldane for an inspirational conversation, and for discussing his
unpublished results (some of which we probably have reproduced). We
also thank John Cardy, Sudip Chakravarty,
Marcel den Nijs and Barry McCoy for helpful
conversations. The work of P.F.\ was supported by a DOE OJI Award, a
Sloan Foundation Fellowship, and by NSF grant DMR-9802813. The work of
C.N. was supported by NSF grant DMR-9983544 and the A.P. Sloan
Foundation.

 \end{document}